\documentclass[a4paper,11pt,oneside]{article}
\usepackage{amsmath,amsfonts,amssymb}
\usepackage{mathrsfs}
\usepackage{graphicx}
\usepackage{bm}
\usepackage{bbm}
\usepackage[affil-it]{authblk}
\usepackage[colorlinks,citecolor=blue,urlcolor=blue,linkcolor=blue]{hyperref}

\begin{document}

\title{Anisotropic Weyl invariance}

\author{Guillem P\'erez-Nadal
	\thanks{E-mail: guillem@df.uba.ar}}
\affil{Departamento de F\'isica, FCEN, Universidad de Buenos Aires\\
Ciudad Universitaria, Pabell\'on 1, 1428 Buenos Aires, Argentina}
\date{June 12, 2017}

\maketitle

\begin{abstract}
We consider a non-relativistic free scalar field theory with a type of anisotropic scale invariance in which the number of coordinates ``scaling like time'' is generically greater than one. We propose the Cartesian product of two curved spaces, the metric of each space being parameterized by the other space, as a notion of curved background to which the theory can be extended. We study this type of geometries, and find a family of extensions of the theory to curved backgrounds in which the anisotropic scale invariance is promoted to a local, Weyl-type symmetry.
\end{abstract}


\section{Introduction}

In condensed matter physics, a {\emph{Lifshitz point}} \cite{PhysRevLett.35.1678,chaikin2000principles} is a critical point in the phase diagram of certain systems at which three phases meet, one of which is spatially inhomogeneous. A Lifshitz point is said to be of type $(m,n)$ if the system has $m+n$ dimensions and the inhomogeneous phase exhibits dependence only on the last $n$ coordinates. To lowest order, and discarding boundary terms, the Landau-Ginzburg free energy of a system with scalar order parameter $\phi$ at an $(m,n)$ Lifshitz point is
\begin{equation}\label{flat}
F[\phi]=\frac{1}{2}\int_{{\mathbb R}^m\times{\mathbb R}^n}\phi\left[-\Delta_1+(-\Delta_2)^z\right]\phi,
\end{equation}
where $\Delta_k$ denotes the Laplacian with respect to the $k$-th variable (we view $\phi$ as depending on two variables, the first in ${\mathbb R}^m$ and the second in ${\mathbb R}^n$) and $z>1$ is a natural number (typically $z=2$). An example of a material with a Lifshitz point in its phase diagram is the magnet MnP \cite{PhysRevLett.44.1692,PhysRevB.24.2780}.

Let us regard the above functional as the Euclidean action of a scalar field theory on ${\mathbb R}^m\times {\mathbb R}^n$. This is clearly a non-relativistic theory, because it is not invariant under the full Euclidean group $E(m+n)$.
Rather, it is only invariant under the subgroup $E(m)\times E(n)$.
The theory has also an extra symmetry which is not part of the Euclidean group: it is invariant under transformations ${\mathbb R}^m\times{\mathbb R}^n\to {\mathbb R}^m\times{\mathbb R}^n$ of the form
\begin{equation}\label{ast}
(x,y)\mapsto (\lambda x,\lambda^{1/z}y),
\end{equation}
where $\lambda$ is an arbitrary positive number. These transformations are called {\emph{anisotropic scale transformations}}, and correspond to a stretching of the first $m$ directions and a different stretching of the remaining $n$ directions.

The particular case $m=1$ of the above theory (with the first and second factors of ${\mathbb R}\times{\mathbb R}^n$ regarded as time and space respectively) has attracted considerable attention in the high-energy physics community over the last years, mainly for two reasons. On the one hand, this model has inspired a theory of gravity, called Ho\v rava-Lifshitz gravity \cite{Horava:2008ih,Horava:2009uw,Blas:2009qj}, which is both phenomenologically viable and power-counting renormalizable (in fact, renormalizability has been shown beyond power counting for a simplified version of the theory called the projectable version \cite{Barvinsky:2015kil}). On the other hand, it has been noticed \cite{Kachru:2008yh} that there is an $(n+2)$-dimensional Riemannian manifold (now called Lifshitz space) whose isometries are precisely the symmetries of this model (namely the symmetries described above with $m=1$). By analogy with the AdS/CFT correspondence, this observation has led to the proposal that field theories with these symmetries may admit a dual description in terms of a gravity theory around Lifshitz space, an idea which  has opened a research area known as Lifshitz holography (see \cite{Taylor:2015glc} for a review).

In this context, there has been interest in extending the theory (\ref{flat}) with $m=1$ to curved backgrounds, in order to describe its coupling to Ho\v{r}ava-Lifshitz gravity and also to model the effect of bulk metric perturbations which do not vanish at the boundary of Lifshitz space. Of course, since the theory is non-relativistic one has to specify what is meant by extending it to curved backgrounds. Typically, this means constructing a scalar field theory on an $(n+1)$-dimensional manifold of the form ${\mathbb R}\times{\cal N}$ (whose first and second factors are regarded as time and space respectively) endowed with a metric, which is invariant under all those diffeomorphisms that preserve the foliation ${\cal F}=\{\{t\}\times{\cal N},\,t\in{\mathbb R}\}$ and which reduces to (\ref{flat}) with $m=1$ in the case where ${\cal N}={\mathbb R}^n$ and the metric is the standard Euclidean metric. It is always possible to choose a gauge (the gauge of vanishing shift vector) in which the foliation $\{{\mathbb R}\times\{q\}, q\in{\cal N}\}$ is orthogonal to ${\cal F}$. After this gauge has been chosen, the residual gauge symmetry is the product of diffeomorphism groups ${\text{Diff}}({\mathbb R})\times{\text{Diff}}({\cal N})$. In the cases $n=z=2$ \cite{Baggio:2011ha,Griffin:2012qx} and $n=z=3$ \cite{Adam:2009gq}, extensions of (\ref{flat}) with $m=1$ to curved backgrounds have been obtained in which the anisotropic scale invariance gets promoted to a local, Weyl-type symmetry, and the latter has been shown to be anomalous at the quantum level. The general structure of this anomaly has been studied in \cite{Arav:2014goa} (see also \cite{Arav:2016xjc,Pal:2016rpz,Gomes:2011di}). 

Here we would like to deal with the action (\ref{flat}) in its most general form, namely without the restriction $m=1$ (and also for $n$ and $z$ arbitrary). This will perhaps give rise to new insights into Ho\v{r}ava-Lifshitz gravity and Lifshitz holography as well as their possible generalizations{\footnote{Theories with $m>1$ may be thought of (among other possibilities) as having multiple time dimensions (see \cite{dorling1970dimensionality,Tegmark:1997jg,craig2009determinism} for a discussion about multiple time dimensions in a relativistic context). In the present non-relativistic context, multiple time dimensions arise quite naturally when one attempts to apply dimensional regularization to standard, $m=1$ Lifshitz theories \cite{Arav:2016akx}. I thank the referee for pointing this out to me.}}. By analogy with relativistic theories, the invariance of (\ref{flat}) under the product of Euclidean groups $E(m)\times E(n)$ strongly suggests what an extension of this theory to curved backgrounds should be: a scalar field theory on the product ${\cal M}\times{\cal N}$ of two manifolds, of dimensions $m$ and $n$ respectively, with a metric on each factor, which is invariant under the product of diffeomorphism groups ${\text{Diff}}({\cal M})\times{\text{Diff}}({\cal N})$ and which reduces to (\ref{flat}) in the case where ${\cal M}={\mathbb R}^m$, ${\cal N}={\mathbb R}^n$ and the metrics are the standard Euclidean metrics. On the other hand, by analogy with relativistic scale invariant theories, the symmetry (\ref{ast}) suggests that (\ref{flat}) may admit an extension to curved backgrounds which is invariant under transformations of the form
\begin{equation}\label{awt}
(g,h)\mapsto(\Omega^2g,\Omega^{2/z}h),
\end{equation}
where $g$ is a metric on ${\cal M}$, $h$ is a metric on ${\cal N}$ and $\Omega$ is a positive function on ${\cal M}\times{\cal N}$. It is natural to call these transformations {\emph{anisotropic Weyl transformations}}. Note that, in order for them to make sense, we must allow the metric on each factor of ${\cal M}\times{\cal N}$ to be parameterized by the other factor, i.e., to be a function of two variables, the first in ${\cal M}$ and the second in ${\cal N}$. In the case $m=1$, ${\cal M}={\mathbb R}$, this background geometry is equivalent to the one usually considered (which has been described in the previous paragraph) in the gauge of vanishing shift vector, and our definition of anisotropic Weyl transformations, Eq.~(\ref{awt}), also reduces to the usual one. Studying this type of geometries for arbitrary values of $m$ and finding anisotropic Weyl invariant extensions of (\ref{flat}) to curved backgrounds will be the purpose of this paper.

The paper is organized as follows: sections \ref{bitensors} and \ref{parameterized} are devoted to geometric considerations, and the tools developed there are used in section \ref{anisotropic} to construct anisotropic Weyl invariant extensions of (\ref{flat}) to curved backgrounds. Throughout this paper we work with Euclidean signature.

\section{Bitensor fields}\label{bitensors}

Physical theories on the product of two manifolds are naturally formulated in terms of bitensor fields. A bitensor field is, roughly speaking, a function
of two points which may carry indices associated with each of the points. 
The notion of a bitensor field was first introduced in \cite{J.L.Synge:1960zz}, and it is frequently used in the context of quantum field theory on curved spacetimes, see e.g.~\cite{allen1986vector,PerezNadal:2009hr}. Here we give a precise definition of bitensor fields and study their properties in an explicitly coordinate-independent manner.

We start introducing bitensors over a pair of vector spaces. Let $V$ and $W$ be finite-dimensional real vector spaces, and let $r,s,t,u\in{\mathbb N}$. A {\emph{bitensor of type $(r,s;t,u)$ over $(V,W)$}} is a multilinear map
\begin{equation}
B:(V^*)^r\times V^s\times (W^*)^t\times W^u\to{\mathbb R},
\end{equation}
where the star indicates dual space. In other words, it is a multilinear map that takes $r$ elements of $V^*$, $s$ elements of $V$, $t$ elements of $W^*$ and $u$ elements of $W$ and produces a real number. Note the similarity with the definition of a tensor. In particular, a bitensor of type $(r,s;0,0)$ is a tensor of type $(r,s)$ over the first vector space, and a bitensor of type $(0,0;t,u)$ is a tensor of type $(t,u)$ over the second vector space. 
If $e=\{e_\mu\}$ and $f=\{f_i\}$ are bases of $V$ and $W$ respectively and $\{e^{*\mu}\}$ and $\{f^{*i}\}$ denote the corresponding dual bases, the numbers
\begin{alignat}{2}\label{comp}
&B{}^{\mu_1\dots\mu_r}{}_{\nu_1\dots\nu_s}{}^{i_1\dots i_t}{}_{j_1\dots j_u}=\nonumber\\
&B(e^{*\mu_1},\dots,e^{*\mu_r},e_{\nu_1},\dots,e_{\nu_s};f^{*i_1},\dots,f^{*i_t},f_{j_1},\dots,f_{j_u})
\end{alignat}
are called the {\emph{components of $B$ in the bases $e$ and $f$}}. Because of the multilinearity property, a bitensor is completely determined by its components in some pair of bases. In particular, the components of $B$ in any other
bases $e'$ and $f'$ are given by
\begin{alignat}{2}\label{transf}
B'{}^{\mu_1\dots\mu_r}{}_{\nu_1\dots\nu_s}{}^{i_1\dots i_t}{}_{j_1\dots j_u}&=(\Lambda^{-1})^{\mu_1}_{\phantom{\mu_1}\alpha_1}\dots (\Lambda^{-1})^{\mu_r}_{\phantom{\mu_r}\alpha_r}\Lambda^{\beta_1}_{\phantom{\beta_1}\nu_1}\dots \Lambda^{\beta_s}_{\phantom{\beta_s}\nu_s}\nonumber\\
&\times  (\Gamma^{-1})^{i_1}_{\phantom{i_1}k_1}\dots (\Gamma^{-1})^{i_t}_{\phantom{i_t}k_t}\Gamma^{l_1}_{\phantom{l_1}j_1}\dots \Gamma^{l_u}_{\phantom{l_u}j_u}\nonumber\\
&\times  B{}^{\alpha_1\dots\alpha_r}{}_{\beta_1\dots\beta_s}{}^{k_1\dots k_t}{}_{l_1\dots l_u},
\end{alignat}
where $\Lambda$ and $\Gamma$ are the change of basis matrices, $e'_\mu=\Lambda^{\nu}_{\phantom{\nu}\mu}e_\nu$ and $f'_i=\Gamma^{j}_{\phantom{j}i}f_j$. Note that the components of a bitensor have four types of indices: upper left, lower left, upper right and lower right. Left indices are associated with the first vector space, and right indices are associated with the second vector space. We adopt the convention of denoting left indices with Greek letters and right indices with Latin letters.

There are five operations one can perform with arrays of components of bitensors which are compatible with the transformation law (\ref{transf}), namely which 
correspond to true bitensor operations in the sense that they 
produce the same bitensor independently of the bases chosen: sum of arrays of the same type, product by real numbers, outer product, contraction of an upper left (right) index with a lower left (right) index and permutation of indices of the same type. 
These are the same operations we are used to perform with arrays of components of tensors, with the extra restriction that left indices cannot be contracted or permuted with right indices (note that these forbidden operations are not even defined if the two vector spaces have different dimensions; if they have the same dimension the operations are defined, but they are not compatible with the transformation law (\ref{transf})).
The operations of sum and product by real numbers give the set of all bitensors of type $(r,s;t,u)$ over $(V,W)$ the structure of a vector space, which is easily shown to have dimension $(\dim V)^{r+s}(\dim W)^{t+u}$.

Let ${\cal M}$ and ${\cal N}$ be manifolds. A {\emph{bitensor field of type $(r,s;t,u)$ on ${\cal M}\times{\cal N}$}} is a rule, $B$, that to each pair of points $(p,q)\in{\cal M}\times{\cal N}$ assigns a bitensor, $B(p,q)$, of type $(r,s;t,u)$ over the pair of tangent spaces $(T_p{\cal M},T_q{\cal N})$. 
Bitensor fields are ubiquitous in quantum field theory: the two-point function of a quantum tensor field of type $(r,s)$ is a bitensor field of type $(r,s;r,s)$ on the product of the spacetime manifold with itself.
Suppose that ${\cal M}$ and ${\cal N}$ have dimensions $m$ and $n$ respectively, and let $\varphi:{\cal M}\to{\mathbb R}^m$ and $\psi:{\cal N}\to{\mathbb R}^n$ be coordinate systems. The components of $B$ in the associated bases, expressed as functions of the coordinates (thus viewed as functions ${\mathbb R}^m\times{\mathbb R}^n\to{\mathbb R}$), will be referred to as the {\emph{components of $B$ in the coordinate systems $\varphi$ and $\psi$}}. 
In those cases where we wish to emphasize the pair of coordinate systems chosen, we will denote the array formed by all these components as $B_{\varphi\psi}$. 
A bitensor field is said to be smooth if its components in some (and hence any) pair of coordinate systems are smooth. In what follows, whenever we speak of a bitensor field we will be assuming, if necessary, that it is smooth.

Bitensor fields of type $(r,s;0,0)$ will also be referred to as {\emph{parameterized tensor fields of type $(r,s)$ on the first manifold}}. Note that every such object is a rule that to each pair of points assigns a tensor of type $(r,s)$ at the first point, hence the terminology. For example, the rule that to each pair of points $(p,q)$ in a Riemannian manifold ${\cal M}$ assigns the unit vector at $p$ tangent to the geodesic joining $p$ and $q$ is a parameterized vector field on the first factor of ${\cal M}\times{\cal M}$. As another example, tensor fields on a manifold ${\cal M}$ which depend on an external real parameter (such as the spatial metric in the Hamiltonian formulation of general relativity or, in the context of classical mechanics, the vector field on phase space associated with a time-dependent Hamiltonian) can be viewed as parameterized tensor fields on the first factor of ${\cal M}\times{\mathbb R}$.
Parameterized tensor fields on the first manifold can be differentiated with respect to their second argument (their parameter) in a coordinate-independent way, without need of introducing any extra structure. Indeed, let $F$ be a parameterized tensor field of type $(r,s)$ on the first factor of ${\cal M}\times{\cal N}$. We define the derivative of $F$ with respect to its second argument, $d_2F$, as the bitensor field of type $(r,s;0,1)$ which at each pair of points $(p,q)\in{\cal M}\times{\cal N}$ is given by
\begin{equation}\label{derivative}
(d_2F)(p,q)(v;w)=w(F(p,\phantom{q})(v))
\end{equation}
for any $v\in [(T_p{\cal M})^*]^r\times(T_p{\cal M})^s$ and $w\in T_q{\cal N}$. Note that the expression on the right-hand side above makes sense because $F(p,\phantom{q})(v)$ is a function on ${\cal N}$, and hence we can apply the vector $w$ to it. One can easily check that $(d_2F)(p,q)$ is multilinear, so it is indeed a bitensor at $(p,q)$. The components of $d_2F$ in any pair of coordinate systems are related to those of $F$ by
\begin{equation}\label{derivcomp}
(d_2F)^{\mu_1\dots\mu_r}_{\phantom{\mu_1\dots\mu_r}\nu_1\dots\nu_s i}=\partial_{2i}F^{\mu_1\dots\mu_r}_{\phantom{\mu_1\dots\mu_r}\nu_1\dots\nu_s},
\end{equation}
where $\partial_{2i}$ denotes the partial derivative with respect to the $i$-th coordinate in the second variable (the function $F^{\mu_1\dots\mu_r}_{\phantom{\mu_1\dots\mu_r}\nu_1\dots\nu_s}$ is viewed as depending on two variables, the first in ${\mathbb R}^m$ and the second in ${\mathbb R}^n$). The reason why the derivative of $F$ with respect to its second argument can be defined in a coordinate-independent way without need of introducing any extra structure is easy to understand: the different values that $F(p,q)$ takes as $q$ changes are all tensors at $p$, so they can be compared to each other. Bitensor fields of type $(0,0;t,u)$ will also be referred to as parameterized tensor fields of type $(t,u)$ on the second manifold. The derivative of any such object $S$ with respect to its first argument, $d_1S$, is defined by the obvious analog of (\ref{derivative}), and its components in any pair of coordinate systems are given by the obvious analog of (\ref{derivcomp}).

Let $B$ be a bitensor field of type $(r,s;t,u)$ on ${\cal M}\times{\cal N}$, and let $\sigma:{\cal M}\to{\cal M}$ and $\rho:{\cal N}\to{\cal N}$ be diffeomorphisms. We define the pullback of $B$ by the product $\sigma\times\rho$, denoted $(\sigma\times\rho)_*B$, as the bitensor field of type $(r,s;t,u)$ which at each pair of points $(p,q)\in{\cal M}\times{\cal N}$ is given by 
\begin{equation}
[(\sigma\times\rho)_*B](p,q)(v;w)=B(\sigma(p),\rho(q))(\sigma^*v;\rho^*w)
\end{equation}
for all $v\in[(T_p{\cal M})^*]^r\times(T_p{\cal M})^s$ and $w\in[(T_q{\cal N})^*]^t\times(T_q{\cal N})^u$, where $\sigma^*v$ denotes the tuple of covectors and vectors at $\sigma(p)$ obtained by applying the usual pushforward by $\sigma$ to each element of the tuple $v$, and similarly for $\rho^*w$. Note from the definition of the pullback of a tensor field that
\begin{alignat}{2}
&[(\sigma\times\rho)_*B](p,\phantom{q})(v;\phantom{w})=\rho_*[B(\sigma(p),\phantom{q})(\sigma^*v;\phantom{w})]\label{pullback2}\\
&[(\sigma\times\rho)_*B](\phantom{p},q)(\phantom{v};w)=\sigma_*[B(\phantom{p},\rho(q))(\phantom{v};\rho^*w)].\label{pullback2bis}
\end{alignat}
The expressions on the right-hand side above are ordinary pullbacks of tensor fields. If $\varphi$ and $\psi$ are coordinate systems for $M$ and $N$ respectively, then so are $\varphi\circ\sigma^{-1}$ and $\psi\circ\rho^{-1}$, and it is straightforward to show that the array of components of $(\sigma\times\rho)_*B$ in the coordinate systems $\varphi$ and $\psi$ is given by
\begin{equation}\label{pullbackcomp}
[(\sigma\times\rho)_*B]_{\varphi\psi}=B_{(\varphi\circ\sigma^{-1})(\psi\circ\rho^{-1})}.
\end{equation}
That is, the effect of $(\sigma\times\rho)_*$ on the array of components in a pair of coordinate systems is the same as that of a coordinate transformation. 
From the above equation it readily follows that $(\sigma\times\rho)_*$ acts linearly on the space of all bitensor fields of a given type, preserves the outer product in the sense that $(\sigma\times\rho)_*(A\otimes B)=[(\sigma\times\rho)_*A]\otimes[(\sigma\times\rho)_*B]$ and commutes with contractions and permutations of indices. Particularizing (\ref{pullback2}) to parameterized tensor fields on the first manifold and using it in (\ref{derivative}), we find that
\begin{equation}\label{pullder}
d_2[(\sigma\times\rho)_*F]=(\sigma\times\rho)_*(d_2F)
\end{equation}
for any parameterized tensor field $F$ on the first manifold. In other words, $d_2$ commutes with products of diffeomorphisms. This equation, which can also be shown easily from (\ref{derivcomp}) and (\ref{pullbackcomp}), has an obvious analog for parameterized tensor fields on the second manifold.

\section{Parameterized metrics}\label{parameterized}

We define a {\emph{parameterized metric}} on the first (second) factor of ${\cal M}\times{\cal N}$ as a rule that to each pair of points $(p,q)\in{\cal M}\times{\cal N}$ assigns a metric at $p$ ($q$). Thus, a parameterized metric is a particular case of a parameterized tensor field of type $(0,2)$. An example is the spatial metric in the Hamiltonian formulation of general relativity, whose parameter takes values in ${\mathbb R}$ and is regarded as time.
In this section we study the geometry associated with parameterized metrics.

Let $g$ be a parameterized metric on the first factor of ${\cal M}\times{\cal N}$. For each $q\in{\cal N}$, the tensor field $g(\phantom{p},q):p\in{\cal M}\mapsto g(p,q)$ is a metric on ${\cal M}$, and therefore it has associated a connection, the Levi-Civita connection, which we denote as $\nabla(q)$. We may use this family of connections
(one for each point in ${\cal N}$) to differentiate bitensor fields of any type with respect to their first argument. 
Indeed, let $B$ be a bitensor field of type $(r,s;t,u)$ on ${\cal M}\times{\cal N}$. We define the covariant derivative of $B$ with respect to its first argument, $\nabla_{\!1}B$, as the bitensor field of type $(r,s+1;t,u)$ which at each pair of points $(p,q)\in{\cal M}\times{\cal N}$ is given by
\begin{equation}\label{cov}
(\nabla_{\!1}B)(p,q)(v;w)=\{\nabla(q)[B(\phantom{p},q)(\phantom{v};w)]\}(p)(v)
\end{equation}
for all $v\in [(T_p{\cal M})^*]^r\times (T_p{\cal M})^{s+1}$ and $w\in[(T_q{\cal N})^*]^t\times (T_q{\cal N})^u$.
That is, $(\nabla_{\!1}B)(p,q)(v;w)$ is the covariant derivative of the tensor field $B(\phantom{p},q)(\phantom{v};w)$ according to the metric $g(\phantom{p},q)$, evaluated at $p$ and acting on $v$. Note from this definition that $\nabla_{\!1}g=0$, and that $\nabla_{\!1}$ agrees with $d_1$ (see (\ref{derivative})) when acting on parameterized tensor fields on the second manifold. The components of $\nabla_{\!1}B$ in any pair of coordinate systems are related to those of $B$ by the equation
\begin{alignat}{2}\label{covcomp}
\nabla_{\!1\alpha}B^{\mu_1\dots\mu_r}{}_{\nu_1\dots\nu_s}{}^{i_1\dots i_t}{}_{j_1\dots j_u}&=\partial_{1\alpha}B^{\mu_1\dots\mu_r}{}_{\nu_1\dots\nu_s}{}^{i_1\dots i_t}{}_{j_1\dots j_u}\nonumber\\
&+\sum_{k=1}^r\Gamma^{\mu_k}_{\phantom{\mu_k}\alpha\beta}B^{\mu_1\dots\beta\dots\mu_r}{}_{\nu_1\dots\nu_s}{}^{i_1\dots i_t}{}_{j_1\dots j_u}\nonumber\\
&-\sum_{l=1}^s\Gamma^{\beta}_{\phantom{\beta}\alpha\nu_l}B^{\mu_1\dots\mu_r}{}_{\nu_1\dots\beta\dots\nu_s}{}^{i_1\dots i_t}{}_{j_1\dots j_u},
\end{alignat}
with Christoffel symbols
\begin{equation}\label{christ}
\Gamma^{\mu}_{\phantom{\mu}\alpha\beta}=\frac{1}{2}g^{\mu\nu}(\partial_{1\alpha}g_{\beta\nu}+\partial_{1\beta}g_{\alpha\nu}-\partial_{1\nu}g_{\alpha\beta}).
\end{equation}
Note the similarity between these equations and the standard coordinate expression for the covariant derivative of a tensor field of type $(r,s)$. Given a parameterized metric on the second manifold, the associated covariant derivative of $B$ with respect to its second argument, $\nabla_{\!2}B$, is defined by the obvious analog of (\ref{cov}), and its components in any pair of coordinate systems are given by the obvious analogs of (\ref{covcomp}) and (\ref{christ}).

We view the dependence of $g$ on its second argument (namely its failure from being just an ordinary metric on ${\cal M}$) as a type of curvature. Thus, the curvature of $g$ is characterized by two bitensor fields: the Riemann tensor $R$, which to every pair of points $(p,q)\in{\cal M}\times{\cal N}$ assigns the value at $p$ of the Riemann tensor of $g(\phantom{p},q)$, and the {\emph{outer curvature}}
\begin{equation}
a=\frac{1}{2}d_2g.
\end{equation}
Note that the Riemann tensor is a parameterized tensor field of type $(1,3)$ on the first manifold, and the outer curvature is a bitensor field of type $(0,2;0,1)$. Their components in any pair of coordinate systems are
\begin{equation}
R{}^{\mu}{}_{\nu\alpha\beta}=\partial_{1\alpha}\Gamma{}^{\mu}{}_{\nu\beta}-\partial_{1\beta}\Gamma{}^{\mu}{}_{\nu\alpha}+\Gamma{}^\rho{}_{\nu\beta}\Gamma{}^\mu{}_{\rho\alpha}-\Gamma{}^\rho{}_{\nu\alpha}\Gamma{}^\mu{}_{\rho\beta},
\end{equation}
where the Christoffel symbols are given by (\ref{christ}), and
\begin{equation}\label{acomp}
a_{\mu\nu i}=\frac{1}{2}\partial_{2i}g_{\mu\nu}.
\end{equation}
For example, suppose that there is a pair of coordinate systems in which $g_{\mu\nu}(x,y)=f(y)\delta_{\mu\nu}$, where $f:{\mathbb R}^n\to{\mathbb R}$ is a positive function and $\delta_{\mu\nu}$ is the Kronecker delta. Then the Riemann tensor of $g$ vanishes but its outer curvature does not, $a_{\mu\nu i}=(\partial_i f)\delta_{\mu\nu}/2$. Clearly, a necessary and sufficient condition for both curvatures to vanish is the existence of a pair of coordinate systems in which $g_{\mu\nu}=\delta_{\mu\nu}$. The Riemann tensor and outer curvature of a parameterized metric on the second manifold are defined analogously, and have analogous properties.

The three objects introduced above (covariant derivative, Riemann tensor and outer curvature) transform in a simple way under the replacement of $g$ by $(\sigma\times\rho)_*g$, where $\sigma:{\cal M}\to{\cal M}$ and $\rho:{\cal N}\to{\cal N}$ are diffeomorphisms. Indeed, if $\nabla_{\!1*}$ is the covariant derivative with respect to the first argument associated with $(\sigma\times\rho)_*g$, we have
\begin{equation}\label{pullnabla}
\nabla_{\!1*}[(\sigma\times\rho)_*B]=(\sigma\times\rho)_*\nabla_{\!1}B
\end{equation}
for any bitensor field $B$. On the other hand, if $R_*$ and $a_*$ denote the Riemann tensor and the outer curvature of $(\sigma\times\rho)_*g$, we have
\begin{equation}\label{pullRQ}
R_*=(\sigma\times\rho)_*R\qquad a_*=(\sigma\times\rho)_*a.
\end{equation}
Eq.~(\ref{pullnabla}) and the first equation in (\ref{pullRQ}) can be shown from (\ref{pullback2bis}) and the transformation properties of the Levi-Civita connection and its Riemann tensor under diffeomorphisms. The second equation in (\ref{pullRQ}) is just a particular case of (\ref{pullder}). Analogous relations hold for a parameterized metric on the second manifold.

Let us now see how the above three objects transform under the replacement of $g$ by $\Omega^2g$, where $\Omega$ is a positive function on ${\cal M}\times {\cal N}$. If $\bar\nabla_{\!1}$ is the covariant derivative with respect to the first argument associated with $\Omega^2 g$, we have in terms of components in any pair of bases
\begin{alignat}{2}\label{weylnabla}
\bar\nabla_{\!1\nu} v^\mu =\nabla_{\!1\nu} v^\mu+\left(2\delta{}^{\mu}{}_{(\nu}\nabla_{\!1\alpha)}\ln\Omega-g_{\nu\alpha}\nabla_{\!1}^{\mu}\ln\Omega\right)v^\alpha
\end{alignat}
for any parameterized vector field $v$ on the first manifold, where the subscript $1$ in $\nabla_{\!1}$ (which should not be regarded as an index) is not involved in the symmetrization. On the other hand, if $\bar R$ and $\bar a$ are the Riemann tensor and the outer curvature of $\Omega^2 g$, we have
\begin{alignat}{2}\label{weylRQ}
\bar R^{\mu}_{\phantom{\mu}\nu\alpha\beta}&=R^{\mu}_{\phantom{\mu}\nu\alpha\beta}+2g_{\nu[\alpha}\left(\nabla_{\!1\beta]}\nabla_{\!1}^\mu\ln\Omega-\nabla_{\!1\beta]}\ln\Omega\,\nabla_{\!1}^\mu\ln\Omega\right)+\nonumber\\
&+2\delta^\mu{}_{[\beta}\left(\nabla_{\!1\alpha]}\nabla_{\!1\nu}\ln\Omega-\nabla_{\!1\alpha]}\ln\Omega\,\nabla_{\!1\nu}\ln\Omega\right)+\nonumber\\
& +2g_{\nu[\alpha}\delta^\mu{}_{\beta]}\nabla_{\!1\rho}\ln\Omega\,\nabla_{\!1}^\rho\ln\Omega\nonumber\\
\bar a_{\mu\nu i} &=\Omega^2\left[a_{\mu\nu i}+g_{\mu\nu} (d_2\ln\Omega)_i\right].
\end{alignat}
Eq.~(\ref{weylnabla}) and the first equation in (\ref{weylRQ}) follow immediately from the transformation properties of the Levi-Civita connection and its Riemann tensor under Weyl transformations (see e.g.~appendix D of \cite{wald1984general}). The second equation in (\ref{weylRQ}) is easily obtained from (\ref{acomp}). Again, analogous relations hold for a parameterized metric on the second manifold.

Let now $g$ and $h$ be parameterized metrics on the first and second factors of ${\cal M}\times {\cal N}$ respectively. We define the integral of a function $f$ on ${\cal M}\times{\cal N}$ according to $g$ and $h$ by the equation
\begin{equation}\label{int}
\int_{{\cal M}\times {\cal N}}f=\int_{\varphi({\cal M})\times\psi({\cal N})}\sqrt{\det(g_{\varphi\psi})\det(h_{\varphi\psi})}\,f_{\varphi\psi},
\end{equation}
where $\varphi:{\cal M}\to{\mathbb R}^m$ and $\psi:{\cal N}\to{\mathbb R}^n$ are coordinate systems{\footnote{Recall that we use the subscript $\varphi\psi$ to denote the array of components in the coordinate systems $\varphi$ and $\psi$. In particular, when attached to a function this subscript indicates that the function is written in terms of the coordinates, i.e., $f_{\varphi\psi}=f\circ(\varphi^{-1}\times\psi^{-1}):\varphi({\cal M})\times\psi({\cal N})\to{\mathbb R}$.}}. Note that the integral on the right-hand side above is just the standard integral of a function on $\varphi({\cal M})\times\psi({\cal N})\subseteq{\mathbb R}^m\times{\mathbb R}^n$. One can check that the latter is independent of the coordinate systems chosen, so the above equation indeed defines the integral of $f$ in an unambiguous way.
If $\int_{{\cal M}\times {\cal N}*}$ denotes the integral according to $(\sigma\times\rho)_*g$ and $(\sigma\times\rho)_*h$, where $\sigma:{\cal M}\to {\cal M}$ and $\rho:{\cal N}\to {\cal N}$ are diffeomorphisms, it clearly follows from (\ref{pullbackcomp}) that
\begin{equation}\label{pullint}
\int_{{\cal M}\times {\cal N}*}(\sigma\times\rho)_*f=\int_{{\cal M}\times {\cal N}} f.
\end{equation}
On the other hand, if $\bar{\!\int}_{\!{\cal M}\times {\cal N}}$ denotes the integral according to $\Omega^2g$ and $\Theta^2h$, where $\Omega$ and $\Theta$ are positive functions on ${\cal M}\times {\cal N}$, we have
\begin{equation}\label{weylint}
\,\,\,\bar{\!\!\!\int}_{\!\!\!{\cal M}\times {\cal N}}\,f=\int_{{\cal M}\times {\cal N}}\Omega^m\Theta^n f.
\end{equation}
Finally, using (\ref{covcomp}), (\ref{christ}) and the counterpart of (\ref{acomp}) for parameterized metrics on the second manifold, one can easily prove a sort of Gauss's law, namely that, except for a boundary term,
\begin{equation}\label{gauss}
\int_{{\cal M}\times {\cal N}}(\nabla_{\!1\mu}+b_{\mu})v^\mu=0
\end{equation}
for any parameterized vector field $v$ on the first manifold, where $b_\mu=h^{ij}b_{\mu ij}$ denotes the trace of the outer curvature of $h$, $b_{\mu ij}=(d_1h)_{\mu ij}/2$. An analogous result holds for parameterized vector fields on the second manifold.
In (\ref{int}) and the equations that followed we implicitly assumed that ${\cal M}$ and ${\cal N}$ admit global coordinate systems. This assumption is in fact not necessary. If we drop it, an equation analogous to (\ref{int}) still defines the integral of $f$ over the product of two coordinate domains. The integral over the whole ${\cal M}\times {\cal N}$ can then be defined with the help of an atlas and a partition of unity for each manifold, along the same lines as in the standard integration of functions over a Riemannian manifold (see e.g.~section 16 of \cite{lee2013introduction}). Eqs.~(\ref{pullint}), (\ref{weylint}) and (\ref{gauss}) still hold if ${\cal M}$ or ${\cal N}$ do not admit global coordinate systems.

\section{Anisotropic Weyl invariance}\label{anisotropic}

Finally we proceed to construct anisotropic Weyl invariant extensions of the action (\ref{flat}) to curved backgrounds. Let ${\cal M}$ and ${\cal N}$ be manifolds of dimensions $m$ and $n$ respectively, and let $g$ and $h$ be parameterized metrics on the first and second factors of ${\cal M}\times {\cal N}$ respectively. 
Consider the operator
\begin{equation}\label{op}
{\cal D}={\cal D}_1+\prod_{s=-J}^J {\cal D}_{2,s}
\end{equation}
on the space of scalar fields on ${\cal M}\times{\cal N}$, where $J=(z-1)/2$ ($z$ is the natural number appearing in (\ref{flat})), the $2J+1=z$ factors in the product are ordered with $s$ increasing from left to right and
\begin{alignat}{2}\label{D's}
&{\cal D}_1=-\left[\nabla_{\!1\mu}+\frac{1}{2}\left(1+\frac{2-m}{n}z\right)b_{\mu}\right]\left[\nabla_{\!1}^\mu+\frac{1}{2}\left(1-\frac{2-m}{n}z\right)b^\mu\right]\\
&{\cal D}_{2,s}=-\left[\nabla_{\!2i}+\frac{1}{2}\left(1+\frac{2-n-4s}{mz}\right)a_{i}\right]\left[\nabla_{\!2}^i+\frac{1}{2}\left(1-\frac{2-n+4s}{mz}\right)a^i\right].\nonumber
\end{alignat}
Here, $\nabla_{\!k}$ (with $k=1,2$) denotes the covariant derivative with respect to the $k$-th argument, and $a_i=g^{\mu\nu}a_{\mu\nu i}$ and $b_\mu=h^{ij}b_{\mu ij}$ are the traces of the outer curvatures of $g$ and $h$ respectively. With the operator (\ref{op}) we construct the action of a scalar field theory on ${\cal M}\times{\cal N}$,
\begin{equation}\label{action}
I[g,h,\phi]=\frac{1}{2}\int_{{\cal M}\times {\cal N}}\phi\,{\cal D}\,\phi.
\end{equation}
Since it is constructed out of covariant objects, this action is invariant under the product of diffeomorphism groups ${\text{Diff}}({\cal M})\times{\text{Diff}}({\cal N})$, that is,
\begin{equation}
I[(\sigma\times\rho)_*g,(\sigma\times\rho)_*h,(\sigma\times\rho)_*\phi]=I[g,h,\phi]
\end{equation}
for any two diffeomorphisms $\sigma:{\cal M}\to {\cal M}$ and $\rho:{\cal N}\to {\cal N}$. This can be seen in detail from (\ref{pullnabla}), the second equation in (\ref{pullRQ}), (\ref{pullint}) and the discussion below (\ref{pullbackcomp}), together with the analogs of the first two of these equations for parameterized metrics on the second manifold. Moreover, the above action clearly reduces to (\ref{flat}) if ${\cal M}={\mathbb R}^m$, ${\cal N}={\mathbb R}^n$ and $g$ and $h$ are the standard Euclidean metrics. And finally, $I$ is anisotropic Weyl invariant, that is,
\begin{equation}\label{awi}
I[\Omega^2g,\Omega^{2/z}h,\Omega^w\phi]=I[g,h,\phi]
\end{equation}
for any positive function $\Omega$ on ${\cal M}\times {\cal N}$, where $w=(2-m-n/z)/2$. Indeed, Eq.~(\ref{weylnabla}) and  the second equation in (\ref{weylRQ}), together with their analogs for parameterized metrics on the second manifold, imply
\begin{alignat}{2}\label{hom1}
&\bar {\cal D}_1(\Omega^w\phi)=\Omega^{w-2}{\cal D}_1 \phi\nonumber\\
&\bar {\cal D}_{2,s}(\Omega^{w-1+(2s+1)/z} \phi)=\Omega^{w-1+(2s-1)/z}{\cal D}_{2,s} \phi,
\end{alignat}
where $\bar {\cal D}_1$ and $\bar {\cal D}_{2,s}$ denote the operators obtained after replacing $(g,h)$ by $(\Omega^2 g,\Omega^{2/z}h)$ in (\ref{D's}). From this equation and (\ref{weylint}) one easily obtains (\ref{awi}). Thus, we conclude that the action (\ref{action}) is an anisotropic Weyl invariant extension of (\ref{flat}) to curved backgrounds.

Although the action (\ref{action}) looks complicated, we may obtain its equation of motion in a fairly easy way. Indeed, consider the inner product 
\begin{equation}
\langle \phi_1,\phi_2\rangle=\int_{{\cal M}\times {\cal N}}\phi_1\phi_2
\end{equation}
on the space of functions on ${\cal M}\times {\cal N}$. Given an operator $O$ on this space, the functional $F[\phi]=\langle\phi,O\phi\rangle$ clearly satisfies $\delta F/\delta\phi=(O+O^\dagger)\phi$. Ignoring boundary terms, we obtain from the Gauss's law (\ref{gauss}) and its analog for parameterized vector fields on the second manifold that
\begin{equation}\label{adjoints}
{\cal D}_{1}^\dagger={\cal D}_1\qquad {\cal D}_{2,s}^\dagger={\cal D}_{2,-s},
\end{equation}
which implies that the operator (\ref{op}) is Hermitian. In consequence, the equation of motion corresponding to the action (\ref{action}) is
\begin{equation}
{\cal D}\phi=0.
\end{equation}
Note that, strictly speaking, the action (\ref{action}) is not functionally differentiable and should be supplemented with a boundary term in order to cancel the boundary terms we ignored in (\ref{adjoints}).

We may in fact construct a large family of anisotropic Weyl invariant extensions of (\ref{flat}) to curved backgrounds. Indeed, consider the functions
\begin{alignat}{2}\label{F's}
&F_1=b_{\mu ij}b^{\mu ij}-\frac{1}{n}b_{\mu}b^\mu \nonumber\\
&F_2=a_{\mu\nu i}a^{\mu\nu i}-\frac{1}{m}a_{i}a^i\nonumber\\
&G_1=R+2\frac{m-1}{n}z\nabla_{\!1\mu}b^\mu-\frac{(m-1)(m-2)}{n^2}z^2b_{\mu}b^\mu\nonumber\\
&G_2=S+2\frac{n-1}{mz}\nabla_{\!2i}a^i-\frac{(n-1)(n-2)}{m^2z^2}a_{i}a^i,
\end{alignat}
where $R$ and $S$ denote the Ricci scalars of $g$ and $h$ respectively. Clearly, these functions transform covariantly under the replacement of $g$ and $h$ by $(\sigma\times\rho)_*g$ and $(\sigma\times\rho)_*h$, where $\sigma:{\cal M}\to{\cal M}$ and $\rho:{\cal N}\to{\cal N}$ are diffeomorphisms. Moreover, they all vanish in the case where ${\cal M}={\mathbb R}^m$, ${\cal N}={\mathbb R}^n$ and $g$ and $h$ are the standard Euclidean metrics. Finally, from the transformation laws (\ref{weylnabla}) and (\ref{weylRQ}), together with their analogs for parameterized metrics on the second manifold, we find 
\begin{alignat}{2}
\bar F_1=\Omega^{-2}F_1\qquad \bar F_2=\Omega^{-2/z}F_2\qquad\bar G_1=\Omega^{-2}G_1\qquad \bar G_2=\Omega^{-2/z}G_2
\end{alignat}
for any positive function $\Omega$ on ${\cal M}\times{\cal N}$, where $\bar F_k$ and $\bar G_k$ denote the functions obtained after replacing $(g,h)$ by $(\Omega^{2}g_1, \Omega^{2/z}g_2)$ in (\ref{F's}). These three properties imply that, if we redefine ${\cal D}$ as the operator (\ref{op}) plus (i) a linear combination of $F_1$ and $G_1$ and/or (ii) the product of operators appearing in (\ref{op}) with any of its factors replaced by a linear combination of $F_2$ and $G_2$, the resulting action (\ref{action}) will still be an anisotropic Weyl invariant extension of (\ref{flat}) to curved backgrounds.

In order to make contact with previous treatments of the subject, let us consider the case $m=1$, $n=z=2$. 
Using (\ref{gauss}) and the second equation in (\ref{adjoints}), we find that in this case the action (\ref{action}) can be rewritten in terms of a pair of coordinate systems as
\begin{alignat}{2}\label{action1}
I[g,h,\phi]=\frac{1}{2}\int_{{\mathbb R}\times{\mathbb R}^2}N\sqrt{\det h}\left[\frac{1}{N^2}\dot\phi^2+(\nabla_{i}\nabla^i\phi)^2\right],
\end{alignat}
where $N$ is the square root of the single component of $g$, the dot indicates partial derivative with respect to the first variable and we have dropped the subscript $2$ in $\nabla_{\!2}$. Regarding the functions (\ref{F's}), note that $F_2$ and $G_1$ vanish for $m=1$. The remaining two functions read in the present case
\begin{eqnarray}\label{F's1}
F_1=K_{ij}K^{ij}-\frac{1}{2}K^2\qquad G_2=S+\frac{\nabla_i\nabla^iN}{N}-\frac{\nabla_iN\nabla^iN}{N^2},
\end{eqnarray}
where $K_{ij}=\dot h_{ij}/(2N)$ and $K=h^{ij}K_{ij}$. The action (\ref{action1}) is the anisotropic Weyl invariant action constructed in \cite{Baggio:2011ha,Griffin:2012qx}, in the gauge of vanishing shift vector. In these references it was also noticed that the scalars (\ref{F's1}) transform homogeneously under anisotropic Weyl transformations. Our equations (\ref{op})-(\ref{action}) and (\ref{F's}) give a remarkable generalization of these results.

\section*{Acknowledgements}
It is a pleasure to thank Gaston Giribet and Diego Blas for discussions and a careful reading of the manuscript. This work has been supported by Fundaci\'on Bunge y Born, UBA and CONICET.

\bibliography{bibweyl}{}
\bibliographystyle{JHEP}

\end{document}